\documentclass[aps,prb,twocolumn,groupedaddress,showpacs]{revtex4}
\usepackage{epsfig}
\begin{document}

\title{Quantum Monte Carlo Study of an  
Interaction-Driven Band Insulator to Metal Transition}
\author{N. Paris$^1$, K. Bouadim$^2$, F. Hebert$^2$,
G.G. Batrouni$^2$, and R.T. Scalettar$^1$}
\affiliation{$^1$Physics Department, University of California, 
Davis, California 95616, USA}
\affiliation{$^2$Institut Non-Lin\'eaire de Nice, UMR 6618 CNRS,
Universit\'e de Nice--Sophia Antipolis, 1361 route des Lucioles, 
06560 Valbonne, France}

\begin{abstract}
We study the transitions from band insulator to metal to Mott
insulator in the ionic Hubbard model on a two dimensional square
lattice using determinant Quantum Monte Carlo.  Evaluation of the
temperature dependence of the conductivity demonstrates that the
metallic region extends for a finite range of interaction values.  The
Mott phase at strong coupling is accompanied by antiferromagnetic (AF)
order.  Inclusion of these intersite correlations changes the phase
diagram qualitatively compared to dynamical mean field theory.
\end{abstract}

\maketitle

\section*{Introduction}

Interaction effects in tight-binding models such as the Hubbard
Hamiltonian have been widely studied, and understood, for their
ability to drive transitions to magnetically ordered states and
insulating behavior.  Also extensively studied, but less well
understood, is the converse phenomenon, namely the possibility that
correlations can cause metallic behavior.  For this latter problem,
attention has focused on an Anderson insulating starting point, in
which the electrons are localized by randomness.  The primitive
picture is that, especially away from commensurate fillings where
correlations tend to localize particles in Mott insulating (MI)
states, electron-electron repulsion can spread out the wave function
and cause delocalization.  Despite considerable effort, whether this
effect actually occurs for fermionic particles in two dimensions, that
is, whether metallic phases exist for disordered, interacting 2d
electron systems, is not settled \cite{2dmitrev}.  In an attempt to
gain leverage on the problem, the phases of disordered and interacting
bosonic systems have now been studied.  Models such as the boson
Hubbard Hamiltonian are more amenable to exact numerical studies and
much is now known about the insulating, glassy, and superfluid phases,
and transitions between them~\cite{bosons}.

A somewhat simpler context in which to study the possibility of
interaction driven insulator to metal transitions is to begin with a
band insulating (BI) state, in which the insulating behavior is caused
by a periodic external potential as opposed to a random one
\cite{bosonsuperlattice,rousseau06}.  Recently, this issue has been
addressed within dynamical mean field theory (DMFT) and a number of
interesting conclusions emerged\cite{garg06}.  However, because DMFT
treats only a single site (retaining, however, all the dynamical
fluctuations of the self-energy which is ignored in conventional,
static mean field theory) it is important to undertake complementary
work which is able to retain intersite fluctuations.

In this paper, we investigate such BI-metal transitions with
determinant Quantum Monte Carlo (DQMC).  
We study the ``ionic Hubbard model":
\begin{eqnarray}
  \label{Hamiltonian}
\hat\mathcal H=&-&t\sum_{\langle lj\rangle \sigma}
       (c_{j\sigma}^\dagger c_{l \sigma} + c_{l\sigma}^\dagger c_{j \sigma})
+U \sum_{l}  n_{l\uparrow}  n_{l\downarrow} 
\nonumber \\
&+& \sum_{l} (\Delta(-1)^l - \mu) (n_{l \uparrow} + n_{l \downarrow})
\,\,\,,
\end{eqnarray}
where $c_{l \sigma}^{\dagger} (c_{l \sigma})$ are the usual fermion
creation (destruction) operators for spin $\sigma$ on site $l$, and
$n_{l \sigma} = c_{l \sigma}^{\dagger} c_{l \sigma}$ is the number
operator.  $t$, $\mu$ and $U$ are the electron hopping, chemical
potential, and on-site interaction strength, respectively.  The kinetic
energy sum is over near neighbor sites $\langle lj\rangle$ 
on a two dimensional square lattice.  $\Delta(-1)^l$ is a
staggered site energy.  In the noninteracting limit, $U=0$, the effect
of $\Delta$ is to produce a dispersion relation, $E(k)=\pm
\sqrt{\epsilon(k)^2 + \Delta^2}$ with $\epsilon(k) = -2t [{\rm cos} k_x
+ {\rm cos}k_y]$, which is gapped at half-filling.  A considerable
amount is known concerning this model in one dimension\cite{onedionichm},
but the existence of an interaction driven metallic phase at
half-filling is still unresolved even in $d=1$.  Metal-insulator
transitions in a related system with {\it randomly} located site
energies with a bimodal distribution have also been studied within DMFT
\cite{byczuk03,byczuk04}.

In this letter we will use DQMC to study the role of interactions in
driving a BI-metal transition in the model described by
Eq.~(\ref{Hamiltonian}).

\section*{Computational Methods}

DQMC \cite{blankenbecler81} provides an exact numerical approach to
study tight binding Hamiltonians like the Hubbard model.  The
partition function $Z$ is first expressed as a path integral by
discretizing the inverse temperature $\beta$. The on-site interaction
is then replaced by a sum over a discrete Hubbard-Stratonovich field
\cite{hirsch85}.  The resulting quadratic form in the fermion
operators can be integrated analytically leaving an expression for $Z$
in terms of a sum over all configurations of the Hubbard-Stratonovich
field with a summand (Boltzmann weight) which is the product of the
determinants of two matrices (one for spin up and one for spin down).
The sum is sampled stochastically using the Metropolis algorithm.  The
results capture correlations in the Hubbard Hamiltonian exactly since
the systematic `Trotter errors' associated with the discretization of
the inverse temperature can easily be extrapolated to zero.  Results
must also be extrapolated to the thermodynamic limit, as we shall
discuss \cite{caveat}.

Equal time operators such as the density and energy are measured by
accumulating appropriate elements, and products of 
elements, of the inverse of the matrix whose determinant gives the
Boltzmann weight.  We will show results for the spin structure factor, 
\begin{eqnarray}
S({\bf k}) =  \sum_{l}e^{i{\bf k} \cdot {\bf l}} \,\, \langle \,\,  
(n_{j+l \uparrow } - n_{j+l \downarrow } ) 
(n_{j \uparrow} - n_{j \downarrow} ) \,\, \rangle \,\, ,
\nonumber
\end{eqnarray}
which probes magnetic order.  For the conductivity, $\sigma_{\rm dc}$,
We employ an approximate procedure\cite{trivedi96} which allows
$\sigma_{\rm dc}$ to be computed from the wavevector ${\bf q}$- and
imaginary time $\tau$-dependent current-current correlation function
$\Lambda_{xx} ({\bf q},\tau)$ without the necessity of performing an
analytic continuation \cite{scalapino93},
\begin{eqnarray}
 \sigma_{\rm dc} = 
   \frac{\beta^2}{\pi} \Lambda_{xx} ({\bf q}=0,\tau=\beta/2) ~.
\nonumber
 \label{eq:condform}
\end{eqnarray}
Here $\beta = 1/T$, $\Lambda_{xx} ({\bf q},\tau) = \langle j_x ({\bf
q},\tau) \, j_x (-{\bf q}, 0) \rangle$, and $j_x ({\bf q},\tau)$ the
${\bf q},\tau$-dependent current in the $x$-direction, is the Fourier
transform of,
\begin{eqnarray}
j_x ({\bf \ell}, \tau) = i \, \sum_\sigma \, t_{{\bf \ell} +
  \hat{x},{\bf \ell}}  \,\,
e^{\tau H}
(c^{\dagger}_{{\bf \ell} + \hat{x},\sigma}
  c^{\phantom \dagger}_{{\bf \ell}\sigma} - c^{\dagger}_{{\bf
  \ell}\sigma} c^{\phantom \dagger}_{{\bf
  \ell}+\hat{x},\sigma})e^{-\tau H} ~.  \nonumber
\end{eqnarray}
This approach has been extensively tested and used for the
superconducting-insulator transition in the attractive Hubbard model
\cite{trivedi96}, as well as for metal-insulator transitions in the
repulsive model \cite{denteneer99_01}.

\section*{Results}

\begin{figure}[t]
\centerline{\epsfig{figure=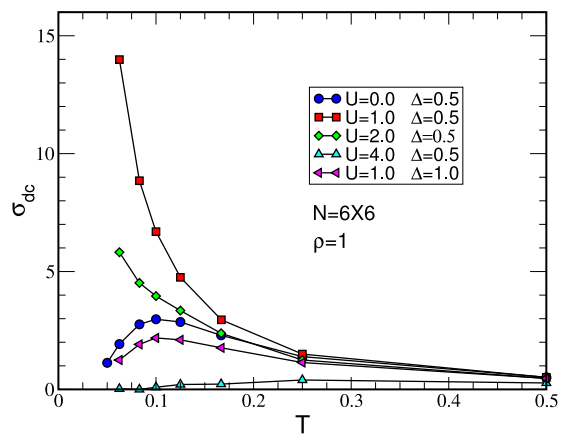,angle=-90,width=7.5cm,clip}}
\caption{The transitions, at half-filling, from a band insulator to
metal to MI with increasing $U$ are shown for periodic potential
strength $\Delta=0.5$.  At $U$=0 the conductivity $\sigma_{\rm dc}$
goes to zero as $T$ is lowered.  However, for $U=1t,2t$ the system is
metallic.  Mott insulating behavior sets in for $U=4t$.  The lattice
size is $6\times 6$.  When $\Delta=1.0$, the band gap increases and
$U=1t$ is no longer sufficiently large to screen the one body
potential and drive the system metallic.  }
\label{sigmavsT}
\end{figure}

We begin by showing the temperature dependence of the conductivity
$\sigma_{\rm dc}$ for increasing values of the interaction strength
for $\Delta=0.5$.  In Fig.~\ref{sigmavsT} we see that the insulating
behavior at $U=0$, signaled by $d\sigma_{\rm dc}/dT > 0$ at low $T$,
is changed to metallic $d\sigma_{\rm dc}/dT < 0$ at low $T$ when
$U=1$.  A further increase of the correlations to $U=2$ weakens the
metallic behavior, which is finally destroyed completely in a
transition to a MI at $U=4$.  When the band gap is larger
($\Delta=1$), the screening of the one-body potential is not
sufficiently strong for $U=1$ to cause metallic behavior, as is shown
by the corresponding data set in Fig.~\ref{sigmavsT}.  Unless
otherwise mentioned, the lattice size used in the simulations is
$N=6\times 6$ and the filling is $\rho=1.0$ (half-filling).

\begin{figure}[t]
\centerline{\epsfig{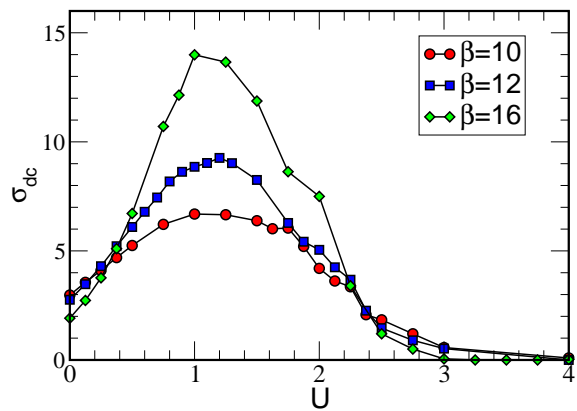}}
\caption{ The conductivity $\sigma_{\rm dc}$ at half-filling for
$\Delta=0.5$ is shown as a function of $U$ for three different low
temperatures, $\beta=10, 12, 16$.  The band-insulator to metal
transition is signaled by the crossing of the curves at $U_{c1}
\approx 0.4t$.  At $U_{c2} \approx 2.4t$ the three curves cross again,
indicating the MI transition.}
\label{sigmavsU}
\end{figure}

In the single site ($t=0$) limit, the ionic Hubbard model is a band
insulator for $U < 2 \Delta$ and a MI for $U > 2 \Delta$.  That is, at
weak coupling and half-filling, the sites with lower energy $-\Delta$
are doubly occupied and those with higher energy $+\Delta$ are empty,
with a gap to further addition of particle set by $2\Delta-U$.  At
strong coupling, both types of sites are singly occupied, with a `Mott' gap
to further addition of particles set by $U-2\Delta$.  At the single
special value $U=2\Delta$ correlations close the gaps
\cite{garg06,rousseau06}.  Figure~\ref{sigmavsU}, which presents
results for $\sigma_{\rm dc}$ for $\Delta=0.5$, shows that when $t$ is
nonzero, this single metallic point is expanded to a finite range of
$U$ values.  Interestingly, however, the largest conductivity remains
near $U=2 \Delta=1$ as one might expect from the $t=0$ analysis.  The BI
to metal transition occurs at $U_{c1} \approx 0.4t$, where the change
in the order of the three curves indicates a transition from
$\sigma_{\rm dc}$ decreasing as $\beta$ increases to $\sigma_{\rm dc}$
increasing as $\beta$ increases.  The metal to MI transition is at
$U_{c2} \approx 2.4t$, where $\sigma_{\rm dc}$ once again decreases as
$\beta$ increases.

\begin{figure}[t]
\centerline{\epsfig{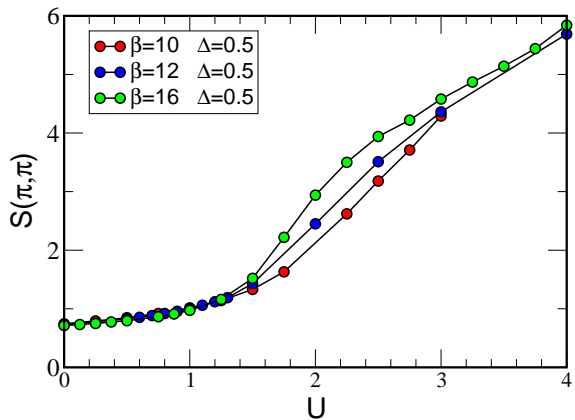}}
\caption{The AF structure factor is shown at half-filling
as a function of $U$ for $\Delta=0.5$ and $\beta=10,12,16$.}
\label{spipi}
\end{figure}

\begin{figure}[t]
\centerline{\epsfig{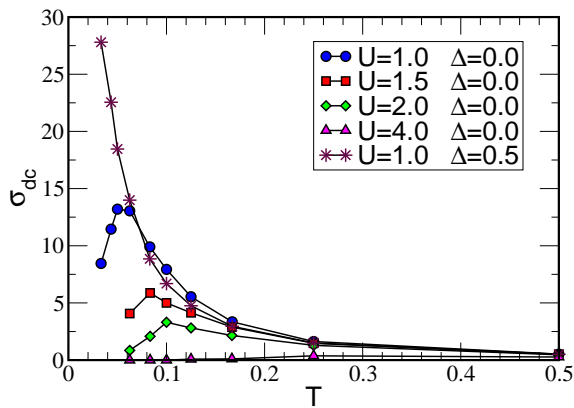}}
\caption{The conductivity $\sigma_{\rm dc}$ is shown as a function of
temperature at half-filling.  When the periodic potential, and hence
the non-interacting band gap, is absent ($\Delta=0.0$) the square
lattice Hubbard model is insulating for {\it all} $U$, due to nesting
of the Fermi surface.  We re-display data for $\Delta=0.5, U=1$ from
Fig.~\ref{sigmavsT} to emphasize the contrast between the metallic
behavior there and the insulating behavior for all $U$ when
$\Delta=0$.  }
\label{sigmavsTD0}
\end{figure}

The use of DQMC to study the ionic Hubbard model allows us to examine
the behavior of intersite correlations, such as the spin-spin
correlations and their Fourier tranform $S({\bf k})$.
Fig.~\ref{spipi} shows results for the AF structure
factor $S(\pi,\pi)$ as a function of $U$ for $\beta=10,12,16$.
Comparing with Fig.~\ref{sigmavsU} we see that the band insulating and
metallic phases are paramagnetic, but that the transition to MI
behavior is accompanied by the onset of AF order.

\begin{figure}[h]
\centerline{\epsfig{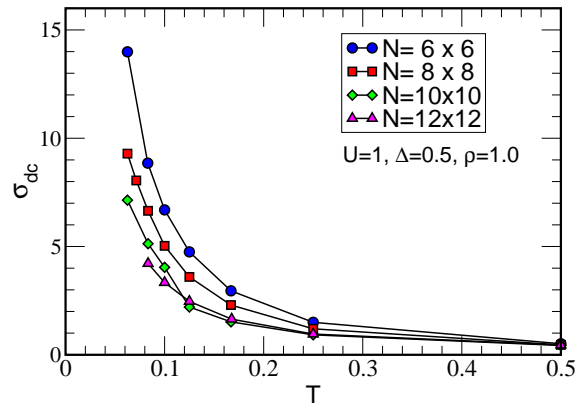}}
\caption{The conductivity at half-filling is shown for different
lattice sizes for $U=1$, close to the point where the system is most
metallic for periodic potential $\Delta=0.5$.  (See
Fig.~\ref{sigmavsU}.) Although $\sigma_{\rm dc}$ decreases with
increasing lattice sizes, the signature of metallic behavior
$(d\sigma_{\rm dc}/dT < 0)$ is unchanged.  }
\label{fss}
\end{figure}

One way in which the inclusion of such intersite correlations changes
the physics in a fundamental way is when the periodic potential is
absent, that is, at $\Delta=0$.  In DMFT in the paramagnetic phase,
the Hubbard model is a metal at weak coupling
\cite{georges96,pruschke95}.  However it is well known that the $d=2$
half-filled square lattice Hubbard model, Eq.~\ref{Hamiltonian}, is an
AF insulator at {\it all} $U$, even weak coupling.  Figure
\ref{sigmavsTD0} presents our results for the conductivity which
confirm this.  At all $U$ values shown, $\sigma_{\rm dc}$ ultimately
decreases as $T$ is lowered.  Indeed, we have verified that the value
of $T$ where $\sigma_{\rm dc}$ has its maximum correlates well with
the temperature $T_*$ at which AF correlations begin to rise rapidly.
This temperature, like the Ne\'el temperature in the $d=3$ Hubbard
model, is a non-monotonic function of $U$, falling to small values
both at weak ($T_* \propto t \, {\rm exp}(-a \sqrt{t/U}$) and at
strong ($T_* \propto t^2/U$) coupling.  To our knowledge, this is the
first time the insulating nature of the square lattice Hubbard model
at weak coupling has been shown from Quantum Monte Carlo studies of
$\sigma_{\rm dc}$.  It is interesting to note that while all the
$\Delta=0$ curves share a common low temperature slope $d\sigma_{\rm
dc}/dT > 0$, a distinction between the origins of insulating behavior
in the AF and Mott is clearly evident.  At small $U$, $\sigma_{\rm
dc}$ attains a much larger value before turning over as $T$ is lowered
than in the strong coupling Mott regime.

\begin{figure}[h]
\centerline{\epsfig{figure=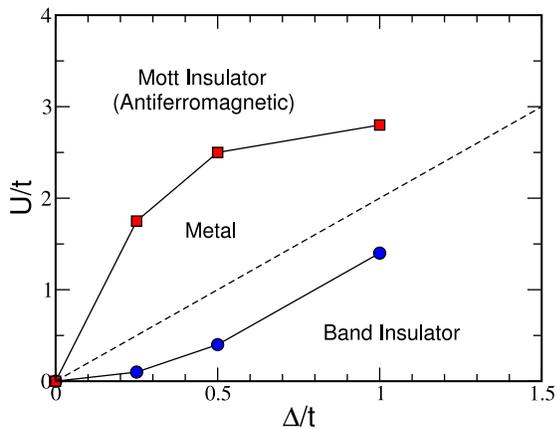,angle=-90,width=7.5cm,clip}}
\caption{The phase diagram of the ionic Hubbard model.
Symbols are the result of our QMC simulations.  The dashed line
is the strong coupling (t=0) phase boundary between band-insulator
and Mott insulator.}
\label{pd}
\end{figure}

While DQMC allows us to look at intersite correlations and concomitant
phenomena like antiferromagnetism, the method
employs lattices of finite size, unlike DMFT which directly probes the
thermodynamic limit.  Thus, it is important to verify that the
metallic phase we observe persists on larger lattices.  In
Fig.~\ref{fss} we show results for $\sigma_{\rm dc}$ as a function of
temperature in the metallic phase for lattices up to $12 \times 12$.
The rise in $\sigma_{\rm dc}$ with decreasing $T$ is seen to occur for
all the lattices studied.  We comment that it is not surprising that
we find the lattice size has a rather substantial influence on the
conductivity for these parameters, since it is known that such finite
size effects are larger at weak coupling.

\section*{Conclusions}

We have presented determinant Quantum Monte Carlo studies of the
two-dimensional ionic Hubbard Hamiltonian which demonstrate that interactions
can drive a band insulator metallic.  This work complements DMFT studies
by including intersite AF correlations which
qualitatively alter the ground state phase diagram.

We have focused most of our results on $\Delta=0.5$.  However, we have
also performed simulations sweeping $U$ at $\Delta=0.25$ and
$\Delta=1.00$.  The emerging phase diagram is shown in Fig.~6.  There
are several key differences with that obtained with DMFT
\cite{garg06}.  First, as we have emphasized, the behavior along the
$\Delta=0$ axis is significantly altered.  Contrary to DMFT, the
inclusion of intersite magnetic fluctuations yields an AF insulating
phase for all $U$.  Like the DMFT treatment, however, we find a
metallic phase intervening between band and Mott insulators for
nonzero $\Delta$.  This phase is centered roughly around the strong
coupling boundary (dashed line in Fig.~6).

The sign problem prevents us from performing simulations much beyond
$\Delta/t \approx 1$.  However, we expect that the sign problem will
become better in the limit of large $\Delta$, where we have very
widely separated bands.  Related studies of the boson-Hubbard model in
a ``superlattice'' potential, which exhibit a band-insulator to
superfluid transition \cite{bosonsuperlattice,rousseau06}, show the
appearance of insulating phases at half-integer fillings.  These
`charge-transfer' insulators occur as a result of Mott splitting of
the widely separated bands \cite{byczuk03}.  We plan to explore this
possibility in the fermion case in future work.

We acknowledge support from 
% the National Science Foundation under award
NSF DMR 0312261, and useful input from S.A. Clock.

\noindent
\underbar{Note Added:}  A recent preprint\cite{kancharia06}
reports simulation results for the ionic Hubbard model using cluster
DMFT.  Like our approach, this method incorporates intersite
correlations, and as found here, obtains a Mott phase
along the entire $\Delta=0$ axis of the phase diagram.  A key
difference is that the cluster DMFT approach
suggests the intermediate phase between Mott and band insulators
is a correlated `bond ordered' insulator.


\begin{thebibliography}{10}

\bibitem{2dmitrev} P.A. Lee and T.V. Ramakrishnan,
Rev. Mod. Phys. {\bf 57} , 287 (1985); D. Belitz and
T.R. Kirkpatrick, Rev. Mod. Phys. {\bf 66} , 261 (1994);
``Metallic behavior and related phenomena in two dimensions,''
E. Abrahams, S.~V. Kravchenko, and M.~P. Sarachik,
Rev. Mod. Phys. {\bf 73}, 251 (2001);
S. V. Kravchenko and M. P. Sarachik, Rep.
Prog. Phys. {\bf 67}, 1 (2004).

\bibitem{bosons} T. Giamarchi and H. J. Schultz, Phys. Rev. {\bf B37},
325 (1988); G. G. Batrouni {\it et al} Phys. Rev. Lett. {\bf 66}, 3144
(1991); G. G. Batrouni {\it et al}, Phys. Rev. {\bf B48}, 9628 (1993);
M. Cha {\it et al}, Phys. Rev. {\bf B44}, 546 (1998); J. Smakov and
E. Sorensen, Phys. Rev. Lett. {\bf 95}, 180603 (2005).
  

\bibitem{bosonsuperlattice} This problem is also under current
investigation for bosonic systems.  See, for example, 
D. Jaksch {\it et al}, 
% C. Bruder, J.I. Chirac, C.W. Gardiner, and P. Zoller,
Phys.~Rev.~Lett. {\bf 81}, 3108 (1998);
P. Buonsante and A. Vezzani,
Phys. Rev. A {\bf 70}, 033608 (2004).

\bibitem{rousseau06} V.G. Rousseau {\it et al},
% D.P. Arovas, M. Rigol, F. H\'ebert, G.G. Batrouni, R.T. Scalettar, 
Phys. Rev. {\bf B73}, 174516 (2006).

\bibitem{garg06}
A. Garg, H.R. Krishnamurthy, and M. Randeria,
cond-mat/0511351.

\bibitem{onedionichm}  J. Hubbard and J.B. Torrance, 
Phys. Rev. Lett. {\bf 47}, 1750 (1981);
T. Egami, S. Ishihara and M. Tachiki, 
Science {\bf 261}, 130 (1994);
G. Ortiz and R. Martin,
Phys. Rev. B {\bf 49}, 14202 (1994);
R. Resta and S. Sorella,
Phys. Rev. Lett. {\bf 74}, 4738 (1995) and
Phys. Rev. Lett. {\bf 82}, 370 (1999);
M. Fabrizio, A.O. Gogolin, and A.A. Nersesyan,
Phys. Rev. Lett. {\bf 83}, 2014 (1999);
T. Wilkens and R.M. Martin,
Phys. Rev. B {\bf 63}, 235108 (2001); and
C.D. Batista and A.A. Aligia,
Phys. Rev. Lett. {\bf 92}, 246405 (2004).

\bibitem{byczuk03}
K.~Byczuk, M.~Ulmke, and D.~Vollhardt,
Phys.~Rev.~Lett.~{\bf 90},
196403 (2003).

\bibitem{byczuk04}
K.~Byczuk, W.~Hofstetter, and D.~Vollhardt,
Phys.~Rev.~B {\bf 69},
045112 (2004).

\bibitem{blankenbecler81}
R.~Blankenbecler, R.L.~Sugar, and D.J.~Scalapino,
Phys.~Rev.~D {\bf 24}, 2278 (1981).

\bibitem{hirsch85}
J.E.~Hirsch,
Phys.~Rev.~B {\bf 31}, 4403 (1985).

\bibitem{caveat}  The advantages of DMFT relative to DQMC 
are that DMFT works in the thermodynamic limit and that the
`sign problem' is considerably less severe.

\bibitem{trivedi96}
N.~Trivedi, R.T.~Scalettar, and M.~Randeria,
Phys.~Rev.~B {\bf 54}, 3756 (1996).

\bibitem{scalapino93}
 D.J.~Scalapino, S.R.~White, and S.C.~Zhang,
Phys.~Rev.~B {\bf 47}, 7995 (1993).

\bibitem{denteneer99_01}
P.J.H.~Denteneer, R.T.~Scalettar, and N.~Trivedi,
Phys.~Rev.~Lett.~{\bf 83}, 4610 (1999);
% 
% \bibitem{denteneer01}
% P.J.H.~Denteneer, R.T.~Scalettar, and N.~Trivedi,
Phys.~Rev.~Lett.~{\bf 87}, 146401 (2001).

\bibitem{georges96}
A. Georges {\it et al}, 
% G. Kotliar, W. Krauth, and M.J. Rozenberg,
Rev. Mod. Phys. {\bf 68}, 13 (1996).

\bibitem{pruschke95}
T. Pruschke, M. Jarrell, and J.K. Freericks,
Adv. Phys. {\bf 44}, 187 (1995).

\bibitem{kancharia06}
S.S. Kancharia and E. Dagotto,
cond-mat/0607568.

\end{thebibliography}
\end{document}